\documentclass{article}
\usepackage{spconf,amsmath,graphicx,hyperref, pifont, multirow}

\title{Streaming speech recognition with decoder-only large language models 
and latency optimization}
\name{Genshun Wan$^{\dagger 1}$, Wenhui Zhang$^{\dagger 2}$, Jing-Xuan Zhang$^{\star 3}$, Shifu Xiong$^1$, Jianqing Gao$^2$, Zhongfu Ye$^1$\thanks{
$\dagger$Equal contribution. $\star$Corresponding author.\\
This work was supported by the National Natural Science Foundation of China
(Grant No. 62401348), Fundamental Research Funds for the Central Universities (Grant No. GK202406005), Young Talent Fund of Association for Science and
Technology in Shaanxi, China (Grant No. 20250126).}}

\address{
$^1$University of Science and Technology of China, Hefei, P.R.China \\
$^2$iFLYTEK Research, iFLYTEK Co., Ltd., Hefei, P.R.China \\
$^3$School of Artificial Intelligence and Computer Science, 
Shaanxi Normal University, Xi'an, P.R. China \\
\small{\texttt{gswan@mail.ustc.edu.cn, whzhang9@iflytek.com, jxzhanggg@snnu.edu.cn}}
}
\begin{document}
\ninept
\maketitle
\begin{abstract}

Recent advances have demonstrated the potential of decoder-only large language models (LLMs) for automatic speech recognition (ASR). However, enabling streaming recognition within this framework remains a challenge. In this work, we propose a novel streaming ASR approach that integrates a read/write policy network with monotonic chunkwise attention (MoChA) to dynamically segment speech embeddings. These segments are interleaved with label sequences during training, enabling seamless integration with the LLM. During inference, the audio stream is buffered until the MoChA module triggers a read signal, at which point the buffered segment together with the previous token is fed into the LLM for the next token prediction.
We also introduce a minimal-latency training objective to guide the policy network toward accurate segmentation boundaries. Furthermore, we adopt a joint training strategy in which a non-streaming LLM-ASR model and our streaming model share parameters.
Experiments on the AISHELL-1 and AISHELL-2 Mandarin benchmarks demonstrate that our method consistently outperforms recent streaming ASR baselines, achieving character error rates of 5.1\% and 5.5\%, respectively. The latency optimization results in a 62.5\% reduction in average token generation delay with negligible impact on recognition accuracy.

\end{abstract}
\section{Introduction}
\label{sec:intro}

Streaming automatic speech recognition (ASR), also referred to as online ASR, aims to transcribe speech incrementally in real time. It plays a crucial role in practical applications such as live captioning for online meetings and simultaneous translation. A large body of research has explored streaming ASR within conventional end-to-end frameworks, typically relying on unidirectional or blockwise encoders combined with connectionist temporal classification (CTC)~\cite{10.1145/1143844.1143891}, recurrent neural network transducers (RNN-T)~\cite{6638947}, or Transformer transducers~\cite{9053896}. 
In parallel, researchers have also investigated attention-based encoder–decoder (AED) architectures, which adopt label-synchronous decoding. To enable streaming in AED models, several methods have been proposed, including triggered attention~\cite{8683510}, monotonic attention~\cite{10.5555/3305890.3305974}, monotonic chunkwise attention (MoChA)~\cite{chiu2018monotonic}, and its subsequent extensions~\cite{arivazhagan-etal-2019-monotonic, inaguma2020enhancing}.

With the rapid development of large language models (LLMs), leveraging decoder-only architectures for speech recognition has recently attracted considerable attention~\cite{10.5555/3692070.3693991, 10447553, wang-etal-2024-exploring, fathullah2024prompting}. While LLMs have shown remarkable success in non-streaming ASR scenarios, extending them to streaming recognition remains an open challenge. Tsunoo et al.~\cite{tsunoo24_interspeech} proposed a blockwise streaming LLM-based ASR framework with CTC compression, where the LLM predicts output tokens after each block. Jia et al.\cite{10890853} introduced SpeechLLM-XL, which processes speech and text into fixed-size chunks using a CTC force-alignment model for streaming recognition. BESTOW~\cite{10832146} combines GPT-style and T5-style architectures and implements a wait-k strategy for streaming ASR. Chen et al.~\cite{chen24u_interspeech} proposed text token insertion (TTI) and boundary token insertion (BTI) models for LLM-based ASR with discrete speech tokens, where speech boundaries are first extracted using a hybrid ASR system.
Despite these efforts, existing approaches often rely on CTC or hybrid models to perform forced alignment between speech and text in advance. Such cascaded designs complicate end-to-end optimization. Moreover, methods that generate tokens only after fixed-size audio chunks face inherent limitations in adaptively minimizing token-generation latency during streaming recognition.

In this work, we propose a streaming LLM-based ASR method. Our method employs a read/write policy network based on monotonic chunkwise attention (MoChA), which is used for adaptively segmenting the incoming speech before passing it to the LLM for text prediction. Specifically, the MoChA module monitors speech embeddings frame by frame until a read signal is triggered; at that point, the buffered embeddings and the previous token are fed into the LLM to decode the next token. During both training and inference, the audio and text embeddings are interleaved to enable synchronized processing.
For streaming audio encoding, we adopt context-sensitive chunking~\cite{an22_interspeech} with a speech encoder. The encoder outputs are projected through an adaptor and then provided to the LLM as prompts. Owing to the modularized design of the policy network for audio reading, we further apply a minimal latency training (minLT) loss~\cite{inaguma2020minimum}, which effectively reduces the latency of streaming recognition.
The LLM is trained jointly with the speech encoder, policy network, and adaptor using low-rank adaptation (LoRA)~\cite{hu2022lora} in an end-to-end manner. We also propose parameter sharing and joint optimization between the streaming and non-streaming ASR models. This unified approach simplifies the training pipeline and reduces the overall development cost of ASR systems.

We conduct experiments on two widely used Mandarin corpora, AISHELL-1 and AISHELL-2, as well as an in-house multi-domain dataset. The results demonstrate that our method consistently outperforms recently proposed baseline streaming LLM-ASR systems. Incorporating minimal latency training (minLT) effectively reduces recognition delay while maintaining competitive accuracy. Furthermore, ablation experiments confirm the effectiveness of our unified streaming and non-streaming framework and demonstrate the benefits of leveraging pretrained LLM parameters.

\section{Related Works}


Recent studies on LLM-based ASR focus on enhancing speech recognition by leveraging LLMs’ in-context learning~\cite{10447553}, world knowledge~\cite{wang-etal-2024-exploring}, and instruction-following capabilities~\cite{instruction-following-speech}.  
LLM-based ASR systems typically adopt either discrete~\cite{zhang-etal-2023-speechgpt, chen2024loss} or continuous audio representations~\cite{10447553, wu2023decoder, fathullah2024prompting, 10446898}.  
The discrete-token approach, such as SpeechGPT~\cite{zhang-etal-2023-speechgpt}, employs a speech tokenizer model~\cite{hsu2021hubert,zhang2024speechtokenizer} to quantize speech into discrete units, which are then merged with the LLM vocabulary.  
In contrast, methods using continuous audio embeddings~\cite{fathullah2024prompting} rely on a pretrained audio encoder, with its output representations optionally compressed via strided CNNs~\cite{fathullah2024prompting}, frame pooling~\cite{ma2024embarrassingly}, Q-formers~\cite{10445874}, or CTC-based compressors~\cite{wu2023decoder}. The resulting audio embeddings are projected into the LLM word embedding space to prompt transcription.  
Typically, continuous embeddings are treated as soft prompts that are prepended to the label embeddings before being fed into LLMs. Other works adopt a Flamingo-style~\cite{alayrac2022flamingo} integration, where audio features are injected through cross-attention modules to fuse them with the input embeddings or hidden representations of LLMs~\cite{10832146, radhakrishnan-etal-2023-whispering}.  
Our work investigates streaming ASR under the widely used configuration of LLM-based ASR, which employs continuous speech embeddings as soft prompts. 

\begin{figure}
    \centering
    \includegraphics[width=0.75\linewidth]{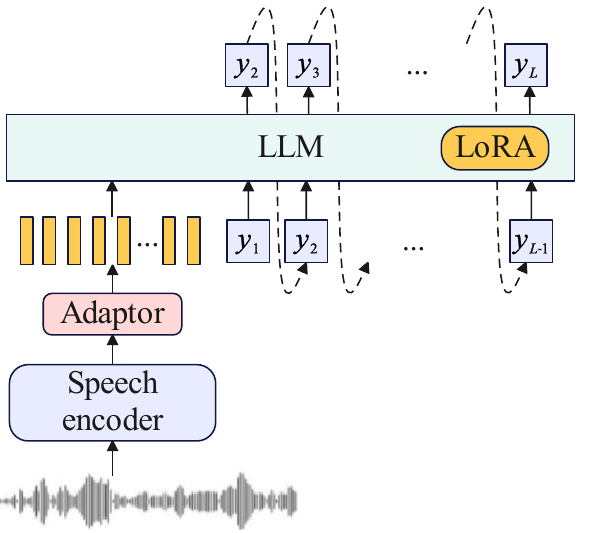}
    \caption{Non-streaming LLM-based ASR architecture. $y_1$ and $y_L$
    represent the BOS and EOS token, respectively.}
    \label{fig:fig1}
\end{figure}

\section{Non-streaming LLM-based ASR}


The architecture of a non-streaming LLM-based ASR model~\cite{fathullah2024prompting, ma2024embarrassingly} is illustrated in Figure~\ref{fig:fig1}.  
The model employs a speech encoder to transform the input speech $X$ into speech representations.  
An adaptor module is then used to project the audio representations into the word embedding space of the LLM.  
Following prior work~\cite{ma2024embarrassingly}, we implement the adaptor as a feed-forward network, which offers both simplicity and competitive performance in ASR tasks.  
The converted audio prompt tokens $h_{1:N}$ are subsequently fed into a decoder-only LLM to generate the corresponding transcription.  
Formally, the conditional probability of the text sequence given the speech input is estimated as
\begin{equation}
    P(Y | X) = \prod_{i=2}^{L} P(y_{i} \mid h_{1:N}, y_{1:i-1}, \theta_{LLM}),
\end{equation}
where $\theta_{LLM}$ denotes the parameters of the LLM.  $y_1, \dots, y_L$
represents the label sequence, where $y_1$ and $y_L$ are special begin of sentence (BOS) and
end of sentence (EOS) token.
The pretrained LLM is commonly equipped with low-rank adaptation (LoRA)~\cite{hu2022lora} weights, enabling memory-efficient finetuning.  
During training, given paired speech–text samples, the model is optimized using a cross-entropy loss to maximize the log-likelihood of the target text sequence.  
The loss is masked over the audio prompt tokens to ensure that only the textual part contributes to optimization.  
At inference time, the input speech is first fully encoded into hidden representations, after which the LLM generates the output text autoregressively until an EOS token is produced.


\begin{figure}
    \centering
    \includegraphics[width=0.76\linewidth]{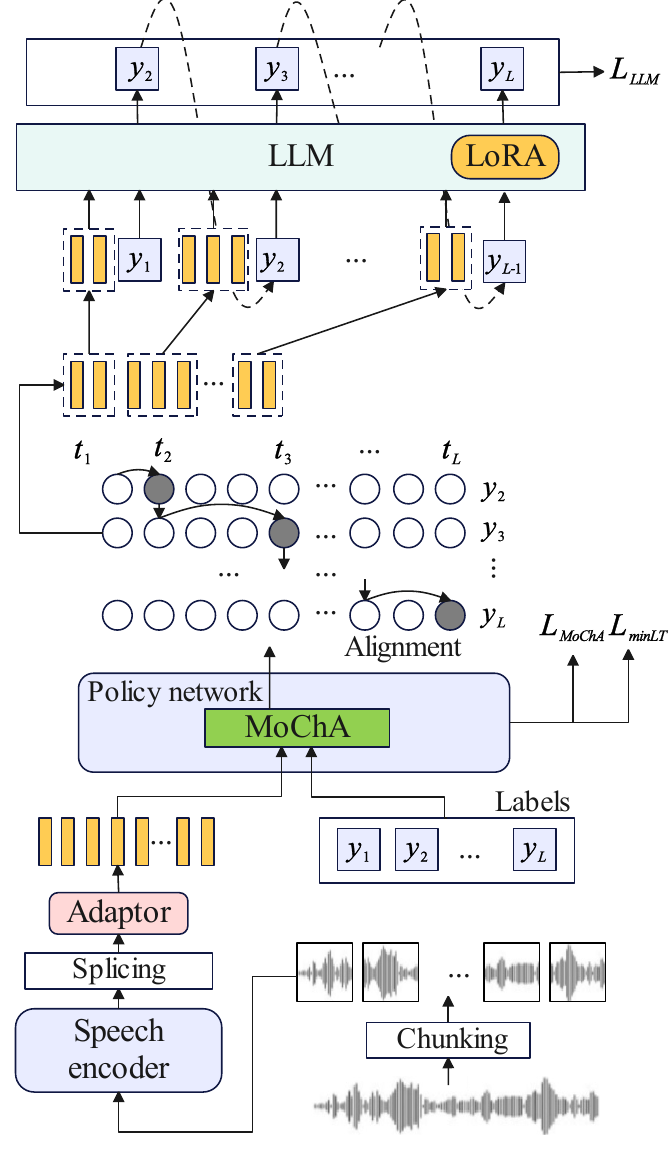}
    \caption{Our proposed streaming LLM-based ASR architecture. $y_1$ and $y_L$
    represent the BOS and EOS token, respectively.}
    \label{fig:fig2}
\end{figure}

\section{Proposed Method}
\subsection{Model architecture}


For streaming speech encoding, we adopt a context-sensitive chunking strategy following prior work~\cite{an22_interspeech}.
The input audio is segmented into chunks, each with an additional history context window. We avoid using any future context to prevent additional encoding latency.
During training, these chunks are processed in parallel by a Conformer-based speech encoder. After discarding the outputs corresponding to the context frames, the remaining chunk-level representations are concatenated to reconstruct the utterance-level audio features.
The resulting speech embeddings are then transformed by an adaptor module into the representation space of the LLM. 
To further align speech and text sequences, we introduce a read/write policy network that adaptively segments the incoming speech stream. This policy enables dynamic synchronization between audio input and text output, as illustrated in Figure~\ref{fig:fig2}.

Our read/write policy network is built upon monotonic chunkwise attention (MoChA)~\cite{chiu2018monotonic}, operating with a lightweight decoder.
At each decoder timestep $i$, the attention mechanism begins scanning the encoder outputs starting from the previously attended position $t_{i-1}$.
A selection probability $p_{i,j}$ is computed for each encoder frame. Once this probability exceeds a predefined threshold at frame $j$, the model triggers a stop-and-decode signal and sets $t_i = j$.
To enhance flexibility, MoChA applies an additional soft chunkwise attention over the hard alignment, allowing the decoder to aggregate information within a local region.
This process generates an encoder index sequence $[t_1, \dots, t_{L}]$ that aligns synchronously with the output tokens $[y_2, \dots, y_L]$, where $t_1=0$.
Intuitively, each token $y_{i}$ for $i=\{2,\dots,L\}$ is decoded based on the acoustic segment $h_{t_{i-1} + 1:t_i} = [h_{t_{i-1} + 1}, \dots, h_{t_i}]$,  which contains the minimal speech context necessary for predicting $y_{i}$.

As shown in Figure~\ref{fig:fig2}, the alignment produced by the policy network is used to interleave segmented speech embeddings with the corresponding text sequence:
\begin{equation}
H_y = [ h_{t_1+1:t_2}, y_1, h_{t_2+1:t_3}, y_2, \dots, h_{t_{L-1}+1:t_{L}}, y_{L-1} ] .
\end{equation}
The resulting mixed sequence $H_y$ is then fed into the LLM during training.
Accordingly, the conditional probability of the text sequence is modeled as:
\begin{equation}
P(Y|X) = \prod_{i=2}^{L} P(y_{i} | h_{t_1+1:t_2}, y_1, \dots, h_{t_{i-1}+1:t_i},  y_{i-1}, \theta_{LLM}) .
\end{equation}
The cross-entropy loss $L_{LLM}$ is only computed at the final frame of each segment, where the next token $y_i$ is predicted.
Our policy network and LLM are trained jointly, enabling dynamic refinement of the segmentation boundaries as training progresses.

\subsection{Training strategy}

The LLM output is optimized with a standard cross-entropy loss $L_{LLM}$.
In addition, the small decoder within the policy network is trained using a cross-entropy loss $L_{MoChA}$, with the same vocabulary as the LLM.
It is important to note that the policy network output is only used during training and discarded at inference.

To further reduce latency, we incorporate a minimal latency training (minLT) loss $L_{minLT}$.
An HMM-based hybrid ASR system is first employed to generate force alignments between speech and text.
Following prior work~\cite{inaguma2020minimum}, we adopt a differentiable expected latency objective:
\begin{equation}
L_{minLT} = \frac{1}{L}\sum_{i=2}^L \sum_{j=1}^N | j \alpha_{i, j} - b_i |,
\end{equation}
where $\alpha_{i,j}$ is the marginalized alignment probability from MoChA, and $b_i$ is the gold boundary from the forced alignment.
The overall training objective is thus:
\begin{equation}
L_{total} = L_{LLM} + L_{MoChA} + \lambda L_{minLT},
\end{equation}
where $\lambda$ is a hyperparameter that balances the latency regularization term.

For efficient fine-tuning of the LLM, we adopt LoRA parameters.
The speech encoder, adaptor, and policy network are trained jointly with LoRA-based optimization from scratch.
Moreover, we propose a joint training scheme that integrates both streaming and non-streaming ASR.
The two models share all parameters but differ in the forward computation path.
During training, each batch is randomly assigned to either the streaming or non-streaming mode, allowing the model to learn both tasks simultaneously.

\subsection{Inference}
During inference, the input speech is first encoded and processed into chunk-level representations.
The policy network then scans these representations until a selection signal is triggered.
Once triggered, the buffered audio segment together with the previous token is passed to the LLM, which generates the next token.
The predicted token is subsequently fed back into both the LLM and the MoChA attention module of the policy network.
This process is repeated iteratively until the LLM outputs an EOS token, completing the transcription.


\section{Experiments}
\subsection{Experiments configuration}
We evaluate our proposed method on three Mandarin datasets: AISHELL-1, AISHELL-2, and a multi-domain in-house dataset (MD).
AISHELL-1~\cite{bu2017aishell} 
 consists of 165 hours of speech (120k/14k/7k utterances for training, development, and testing, respectively).
AISHELL-2\footnote{\url{https://github.com/kaldi-asr/kaldi/blob/master/egs/aishell2}}
 provides 1,000 hours of training data, along with a development set of 2,500 utterances and a test set of 5,000 utterances.
The MD dataset contains roughly 1 hour of speech collected from multiple domains, such as finance, education, and film, and is used exclusively for evaluation.

Our speech encoder is a 12-layer Conformer with 8 attention heads, a hidden dimension of 512, and a feed-forward dimension of 2048. For streaming encoding, we adopt a chunk size of 0.4s, with left context windows of 1.6s. The adaptor is a feed-forward network with a hidden dimension of 1024 and GELU activation.
The LLM is initialized from pretrained Qwen 2.5-1.5B, which consists of 28 Transformer blocks, 12 attention heads, and a hidden dimension of 1536. We retain the original tokenizer and vocabulary to mitigate catastrophic forgetting compared with reinitializing embeddings. Low-rank adaptation (LoRA) weights are applied to the query, key, value, and output projection of the attention modules, with a rank of 32 and scaling factor $\alpha = 64$. The minimal latency (minLT) loss weight $\lambda$ is set to 0.1.

For optimization, we use AdamW with a triangular cyclic learning rate scheduler, which significantly accelerates convergence. The maximum and minimum learning rates are set to $1.5 \times 10^{-4}$ and 0, respectively, with each cycle spanning 25k updates for a total of 100k training steps. During inference, beam search with a beam size of 10 is adopted. 

\subsection{Comparison with baselines}

\begin{table}
   \caption{Character error rates of different methods on the testset of AISHELL-1. $^{\dagger}$Our reproduction results.}
    \label{tab:tab1}
    \centering
    \begin{tabular}{c c c c }
    \hline
    \hline
    Method  & Model type &  Streaming &  CER (\%)  \\
    \hline
     WeNet-U2~\cite{zhang22g_interspeech} & encoder-decoder & \ding{55} & 5.0 \\
     Baseline-non-stream & encoder-decoder &  \ding{55} & 6.5 \\
     Baseline-stream & encoder-decoder & \ding{51} & 6.9 \\
     BTI~\cite{chen24u_interspeech}            & decoder-only   & \ding{51} & 5.9 \\
     BESTOW$^{\dagger}$~\cite{10832146}         & decoder-only    & \ding{51}  & 5.3 \\
     \hline
     \multirow{2}{*}{\emph{Proposed}}     &  \multirow{2}{*}{decoder-only} & \ding{55} & 4.9 \\
                        &   & \ding{51} & 5.1 \\
\hline
\hline
    \end{tabular}
\end{table}

\begin{table}
   \caption{Character error rates of different methods on the testset of AISHELL-2. $^{\dagger}$Our reproduction results.}
    \label{tab:tab2}
    \centering
    \begin{tabular}{c c c c }
    \hline
    \hline
    Method  & Model type &  Streaming &  CER (\%)  \\
    \hline
     WeNet-U2~\cite{zhang22g_interspeech} & encoder-decoder & \ding{55} & 6.1 \\
     Baseline-non-stream & encoder-decoder &  \ding{55} & 5.9 \\
     Baseline-stream & encoder-decoder & \ding{51} & 6.1 \\
     BTI~\cite{chen24u_interspeech}            & decoder-only   & \ding{51} & 7.2 \\
     BESTOW$^{\dagger}$~\cite{10832146}         & decoder-only    & \ding{51}  & 5.6 \\
     \hline
     \multirow{2}{*}{\emph{Proposed}}     &  \multirow{2}{*}{decoder-only} & \ding{55} & 5.0 \\
                        &   & \ding{51} & 5.5 \\
\hline
\hline
    \end{tabular}
\end{table}

\begin{table}[t]
   \caption{Character error rates of different methods on our in-house MD dataset.}
    \label{tab:tab3}
    \centering
    \begin{tabular}{c c c c }
    \hline
    \hline
    Method  & Model type &  Streaming &  CER (\%)  \\
    \hline
     Baseline-non-stream & encoder-decoder &  \ding{55} & 8.0 \\
     Baseline-stream & encoder-decoder & \ding{51} & 9.6 \\
     \hline
     \multirow{2}{*}{\emph{Proposed}}     &  \multirow{2}{*}{decoder-only} & \ding{55} & 6.7 \\
                        &   & \ding{51} & 7.6 \\
\hline
\hline
    \end{tabular}
\end{table}

We construct two encoder–decoder based models as baselines: a non-streaming model (\textbf{Baseline-non-stream}) and a streaming model (\textbf{Baseline-stream}) that employs MoChA attention. Our proposed model is trained on AISHELL-1 and AISHELL-2, and the results are reported in Table~\ref{tab:tab1} and Table~\ref{tab:tab2}, respectively. In addition, we evaluate the AISHELL-2 trained model on our in-house MD dataset (Table~\ref{tab:tab3}).
Unlike the baselines, our method unifies the training of streaming and non-streaming models, enabling evaluation in both modes. As shown in Table~\ref{tab:tab1}–\ref{tab:tab3}, the proposed model achieves the best performance in the non-streaming setting across all datasets, demonstrating the effectiveness of leveraging LLMs for speech recognition. In the streaming setting, the recognition accuracy slightly decreases compared with the non-streaming mode, but our model consistently outperforms the streaming baseline, confirming the effectiveness of the proposed decoder-only LLM framework for streaming ASR.



\subsection{Latency optimization}
This section investigates the effectiveness of the minimal latency (minLT) training loss in optimizing decoding latency. We compute the latency for predicting the first (First), middle (Mid.), and last (Last) tokens using force-alignment results, as shown in Table~\ref{tab:tab4}. We also report the average latency (Avg.) during streaming decoding.
The latency is measured with frames,
where one frame equals to 40ms.
From Table~\ref{tab:tab4}, we observe that introducing the minLT loss significantly reduces the latency of token generation. Meanwhile, the CER increases only marginally from 5.4\% to 5.5\%. These results demonstrate that our method effectively enables streaming ASR with substantially lower token generation latency while maintaining competitive recognition accuracy.


\begin{table}[]
\caption{Character error rates and latency for streaming ASR
on the testset of AISHELL-2. 
Latency is measured in frames, where one frame equals 40 ms.
}
    \label{tab:tab4}
    \centering
    \begin{tabular}{c c c c c c}
    \hline
    \hline
    \multirow{2}{*}{Method}    & \multirow{2}{*}{CER (\%)} & \multicolumn{4}{c}{Latency (frame)} \\
    & & First & Mid. & Last & Avg. \\
    \hline
    Baseline-stream & 6.1 & 19 & 15 & 7 & 15 \\
    \emph{Proposed}-w/o minLT & 5.4 & 18 & 15 & 9 & 16 \\
    \hline
    \emph{Proposed}  & 5.5 & 10 & 5 & 2 & 6 \\
    \hline
    \hline
    \end{tabular}    
\end{table}

\begin{table}[t!]
    \caption{Charater error rates in ablation studies 
    on the testset of AISHELL-2.}
    \label{tab:tab5}
    \centering
    \begin{tabular}{l c c}
    \hline
    \hline
     \multirow{2}{*}{Method} & Non-streaming & Streaming \\
     &  CER (\%) & CER (\%) \\
     \hline
    \emph{Proposed}     & 5.0 & 5.5 \\
    \qquad -w/o joint-train &  5.1 & 5.6 \\
    \qquad -w/o LoRA & 5.4  & 5.7  \\
    \qquad -w/o Qwen init. &  6.5   & 7.2   \\
    \hline
    \hline
    \end{tabular}

\end{table}

\subsection{Ablation studies}
\label{sec:subsecab}
We conduct ablation studies to examine three key design choices: the joint training of streaming and non-streaming models (\textbf{w/o joint-train}), the use of LoRA for fine-tuning (\textbf{w/o LoRA}), and the initialization with pretrained Qwen 2.5 parameters (\textbf{w/o Qwen init.}). The results are summarized in Table~\ref{tab:tab5}.
As shown in the table, training streaming and non-streaming models separately yields performance comparable to our unified training strategy. This indicates that a single unified model can effectively support both modes without performance degradation, thereby simplifying model development.
When LoRA is removed and the pretrained LLM is frozen (w/o LoRA), performance degrades, highlighting the importance of efficient parameter adaptation. Furthermore, when the LLM is randomly initialized instead of using pretrained parameters (w/o Qwen init.), the performance drops significantly. These results confirm the crucial role of leveraging pretrained LLM knowledge for ASR.


\section{Conclusion}

In this work, we proposed a streaming speech recognition method built on a decoder-only large language model (LLM). A policy network based on monotonic chunkwise attention adaptively segments the audio input, which is then decoded by the LLM in a streaming fashion. This design enables end-to-end training and latency optimization through a minimal latency training strategy.
We further introduced a joint training framework for both streaming and non-streaming modes. Experimental results on AISHELL-1, AISHELL-2, and an in-house multi-domain dataset demonstrate that our method consistently outperforms recently proposed streaming LLM-based ASR baselines. In addition, the results confirm the effectiveness of latency optimization and the advantages of unifying streaming and non-streaming models within a single framework.


\vfill\pagebreak





\bibliographystyle{IEEEbib}
\bibliography{strings,refs}

\end{document}